# Double Resonance Nanolaser based on Coupled Slit-hole Resonator Structures

Z. H. Zhu,[1,2]   H. Liu,[1,*]   S. M. Wang,[1]   W. M. Ye,[2]   X. D. Yuan,[2]   S. N. Zhu,[1,*]

[1]*Department of Physics, National Laboratory of Solid State Microstructures, Nanjing University, Nanjing 210093, People's Republic of China*

[2]*College of Opto-Electronic Engineering, National University of Defense Technology, Changsha 410073, People's Republic of China*

This work investigates a kind of metallic magnetic cavity based on slit-hole resonators (SHRs). Two orthogonal hybrid magnetic resonance modes of the cavity with a large spatial overlap are predesigned at the wavelengths of 980 nm and 1550 nm. The Yb-Er co-doped material serving as a gain medium is set in the cavity; this enables the resonator to have high optical activity. The numerical result shows that the strong lasing at 1550 nm may be achieved when the cavity array is pumped at 980 nm. This double resonance nanolaser array has potential applications in future optical devices and quantum information techniques.

Recently, researchers have reported the observation of the stimulated amplification of surface plasmon polaritons (SPP) at different frequencies.[1-10] These remarkable observations provide a compelling starting point for the applications of SPP in optoelectronics integration. Analogous to SPP, the split-ring resonator (SRR), combined with semiconductor materials or quantum-dot-doped dielectrics, is also proposed as an efficient way to produce nanolasers.[11,12] In our recent work, we have proposed and theoretically analyzed a magnetic resonance nanosandwich nanolaser.[13] This structure is for illustration purposes only because the spatial overlap between the pumping mode (higher order mode) and the lasing mode (lower order mode) is small. The other drawback lies in the difficulty to fabricate it; it is simply difficult to create it in experiments.

We then propose a new kind of magnetic resonators, which are slit-hole resonators (SHRs).[14] Compared with SRR, SHR is more easily fabricated in nanoscale. Magnetic resonance can be easily created in infra-red range in such structures. In this work, we propose a new kind of nanolaser based on orthogonal coupled SHR structures (CSHR). The CSHR are carefully designed to provide two magnetic resonance modes (980 nm and 1550 nm) with a large spatial overlap. Combined with the Ytterbium-erbium co-doped gain material, this structure can have resonance not only at the lasing wavelength (1550 nm) but also at the pumping wavelength (980 nm). There are several Er:Yb co-doped gain materials, such as Er:Yb:phosphate glasses, Er:Yb:YAG, Er:Yb:YSO, and Er:Yb:CAS crystals.[15,16] We selected the Er:Yb:YCa$_4$O(BO$_3$)$_3$ (Er:Yb:YCOB) crystal[17] as the gain medium for our principal demonstration because it is widely used in generating light in the optical communication wavelength region at around $1.5\,\mu m$, mainly due to its ease of growth, non-hygroscopicity, good mechanical properties, acceptable thermal conductivity, and large absorption and emission cross section. We then introduce a set of rate equations to describe the operation of the laser and predict the lasing condition.

Figure 1 shows our designed CSHR structure arrays, which are based on the designing idea proposed in Ref. 14. Fig. 1(b) indicates a unit cell of the array [whose cutting plane schematic is shown in Fig. 1(c) and Fig.1 (d)]. It comprises of two parts: the three nanoholes drilled on the silver film and two slits linking the three holes.

The geometry parameters of the CSHR shown are as follows. (1) The thicknesses $h_1$, $h_2$,

and $h_0$ of the silver film, gain medium layer above the silver film, and SiO$_2$ glass substrate are 40 nm, 250 nm, and 800nm, respectively. (2) The diameter D of the nanohole is 85 nm. (3) The distance between two neighbor nanohole centers is 120 nm. (4) The width w of the slit is 40 nm. (5) The distance between the nearest-neighbor single laser cell structures is 1,000 nm. (6) The silver is treated as a dispersive medium following the Drude model. The metal permittivity in the infrared spectral range is derived by using $\varepsilon(\omega) = \varepsilon_\infty - \omega_p^2/(\omega^2 + i\omega/\tau)$. (7) The values of $\varepsilon_\infty, \omega_p$ and $\tau$ fitted to the experimental data in the 950–1800 nm wavelength range are 1.0, $1.38 \times 10^{16}$ rad/s, and 33 fs, respectively. (8) The refractive index of Er:Yb:YCOB and SiO$_2$ substrate in the 950–1800 nm wavelength range, measured through an ellipse spectrometer, are 1.35 and 1.5, respectively.

Based on Pendry's theory,[18] the drilled hole at high frequency can be regarded as an effective inductance, while the slit can be considered an effective capacitance. Therefore, the structure shown in Fig. 1(b) can be seen as an equivalent inductance-capacitance (LC) circuit. When magnetic resonance occurs, the electromagnetic energy is mainly stored in the capacitor (i.e., the slit), and thus the structure can be considered a cavity of confining light. According to the geometric symmetry in the CSHR design, we can expect this structure to possess two magnetic plasmon resonance modes with symmetry and asymmetry. We were also able to predict that the two modes would have a large spatial overlap between their electric fields. This is considered a key to and beneficial to lasing. We used the parallel finite difference time domain (FDTD) method with a perfectly matched layer boundary condition for the calculation of the magnetic resonance to examine the exact mode characters. We placed a pulsed dipole source (power far below threshold) in the asymmetry position to excite the SHR structure and record the fields at the monitor points positioned in different locations in the structure. We obtained the resonant frequency by using the fast Fourier transform (FFT) of electric field at the monitor points. Afterwards, the field was initialized by a dipole oscillation source with the angle frequency equivalent to the resonant frequency. A resonance mode was formed and its mode distribution was obtained in the simulations, while the local fields decayed over time. After the mode was formed, we switched off the dipole oscillation ($t_0$ is the switch-off time) and calculated the electromagnetic energy $U$ (neglecting the field in silver films) as a function of time $t$. Q

($Q = [-\omega(t-t_0)]/\ln[U(t)/U(t_0)]$) was calculated from the slope of the logarithm energy-time relation. The effective mode volume $V_{eff}$ ($V_{eff} = [\int \varepsilon(r)|E(r)|^2 d^3r]/\max[\varepsilon(r)|E(r)|^2]$) was obtained[2,19] based on the ratio of total electric-field energy to the maximum value of electric-field energy density in the structure. The Purcell factor was derived by using $F = 3Q\lambda^3/(4\pi^2 V_{eff} n^3)$,[20] where n and λ are the refractive index of the gain material and the wavelength in vacuum, respectively.

Fig. 2(a) shows the resonant wavelengths obtained through FFT. Fig. 2(a) presents how the design of CSHR forms two resonance modes. The resonant wavelengths are 980 nm and 1550 nm. The electric field patterns for the two modes are calculated and shown in Figs. 2(b) and 2(c). Figs. 2(b) and 2(c) show that the electric field of two resonance modes are both focused inside the two slits. This means that the two modes have a very large spatial overlap, which is very important in producing high laser output efficiency. As the 980 nm and 1550 nm wavelengths are respectively close to the absorption peak and emission peak of the gain material Er:Yb:YCOB, we can select a mode (980 nm) as the pumping mode and another as the lasing mode (1,550 nm).

Fig. 3 shows the energy level diagram for the Er–Yb co-doped system. The operation principle of the CSHR laser is as follows. When pumping light (980 nm) with s-polarization ($\vec{E} \parallel -\hat{x}+\hat{z}$) is applied on the structure [see Fig. 1(a)], the asymmetry magnetic resonance is excited, which results in the light energy being strongly coupled into the two slits [see Fig. 2(c)]. Afterwards, the pumping light is absorbed by $Yb^{3+}$ ions (associated with the $^2F_{7/2} \rightarrow {}^2F_{5/2}$ transition, see Fig. 3). The excitation is transferred to the $Er^{3+}$ ions from the $^2F_{5/2}(Yb^{3+})$ level based on the resonant transferring mechanism: $^2F_{5/2} \rightarrow {}^2F_{7/2}$ ($Yb^{3+}$), $^4I_{15/2} \rightarrow {}^4I_{11/2}$ ($Er^{3+}$). The $^4I_{11/2}$ level has a relatively low lifetime and rapidly relaxes nonradiatively to the $^4I_{13/2}$ level, releasing energy as vibrational energy namely phonons. The $^4I_{13/2}$ level is metastable, possessing a lifetime of around 10 ms. Therefore, the population inversion between the $^4I_{13/2}$ and $^4I_{15/2}$ levels in $Er^{3+}$ becomes possible. Lasing at 1,550 nm wavelength can be achieved through stimulated emission from $^4I_{13/2}$ level to $^4I_{15/2}$ level [see Fig. 3(b)]. Each unit of array structure at the far field can be considered a lasing source. The radiation of waves from these lasing source arrays superposes in free space.

The superposition produces an emitting beam. The polarization of the emitting beam is determined by the characteristic of the symmetry magnetic resonance [Fig. 2(a)] and is p-polarization ($\vec{E} \parallel \hat{x} + \hat{z}$) [see Fig. 1(a)].

We used a set of rate equations to model the operation of the laser to predict the lasing condition. The set of rate equations taking into account spatial distributions of two modes have been clearly described in our previous work.[13] Therefore, we no longer provide the detailed equations and material parameters used in the calculation. Using the FDTD method, we calculated and obtained the quality factor and effective mode volume of the pumping and lasing modes, which are $15, 0.03(\lambda/n)^3$ and $20, 0.005(\lambda/n)^3$, respectively. The Purcell factor of the pumping mode is 40. Using these mode parameters in the equations and assuming the total photon number of lasing mode in the cavity to be 1 and the pumping rate to be high enough, we then derived the $Er^{3+}$ threshold doping concentrations as $8 \times 10^{26} ions/m^3$ (the $Yb^{3+}$ concentration is fixed at $5.0 \times 10^{27} ions/m^3$). Choosing $Er^{3+}$ concentration at $5.0 \times 10^{27} ions/m^3$, we calculated and obtained the corresponding absorbed threshold pumping power ($\hbar\omega_p \sigma_Y v_p F_p N_p N_Y / \eta_p$, $\eta_p$ is the quantum efficiency and approximately equal to 1) as $2.0 \times 10^{-5} W$. The lasing slope efficiency derived from the lasing output power and absorbed pumping power was about 40%. The typical $Er^{3+}$ concentration in Er:Yb co-doped gain materials was in the range of $10^{25} \sim 10^{27}$ ion/$m^3$. Therefore, the coupled slit-hole resonator structure combined with the Er:Yb:YCOB gain materials can theoretically realize lasing. The proposed nanolaser can be easily extrapolated using other gain materials such as the quantum-well or quantum dot material system (e.g., InGaAsP-InP), whose gain is much larger than the Er:Yb:YCOB used in this paper.

In summary, we have proposed a new kind of metallic nanolaser based on coupled slit-hole resonators. Two hybrid magnetic plasmon resonance modes are formed in this structure due to the strong coupling effect. Both the pump light and the lasing wave can resonate simultaneously and have a very large spatial overlap by carefully devising the parameters. This results to a high pumping efficiency and a complete polarization conversion between the pump light and lasing output beam. The proposed structure is simple. Moreover, the metallic film layer can be considered a component of planar substrate, which provides many advantageous features for the practical application to lasers. These features include easy integration in photonic and electronic integration circuits, and ability to act as a heat sink path under intense optical pumping. Outside of its use in nanolaser, the proposed coupled SHR cavities can also be used in many other nonlinear optical processes and quantum information techniques, such as enhanced Raman scattering, bio-sensors, and hyper-entanglement photon sources.

This work is supported by these National Programs of China (Grant Nos. 60907009, 10704036, 10604029, and Grant No. 2006CB921804).

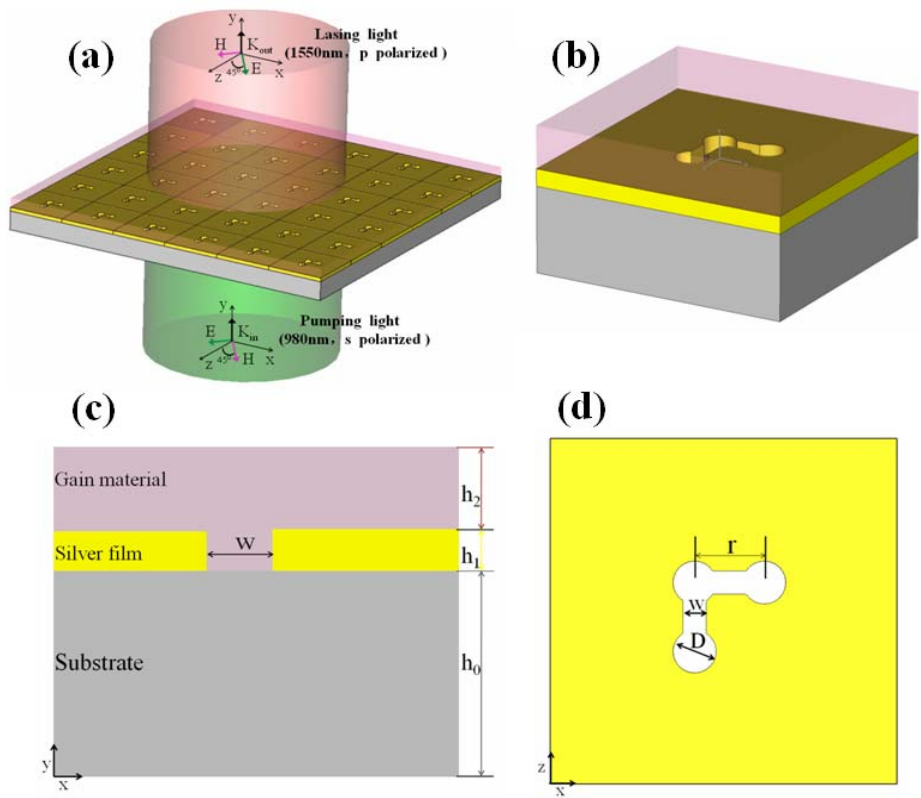

Fig.1 Coupled slit-hole resonator nanolaser structures. (a) Laser arrays. (b) Single laser cell structure. (c) Cutting plane (y-x plane) of single laser cell structure. (d) Cutting plane (z-x plane) of single laser cell structure.

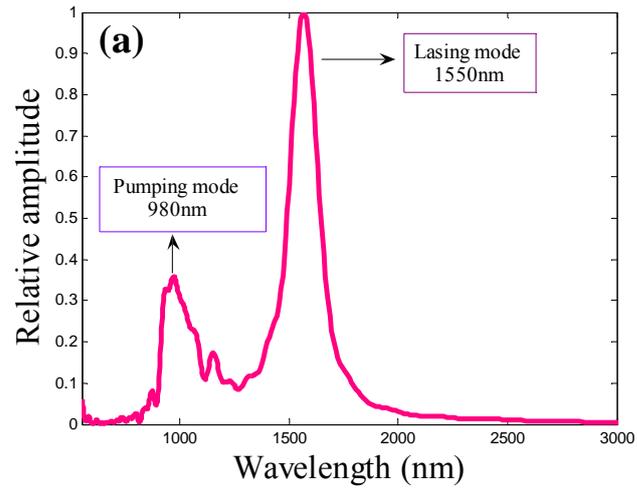

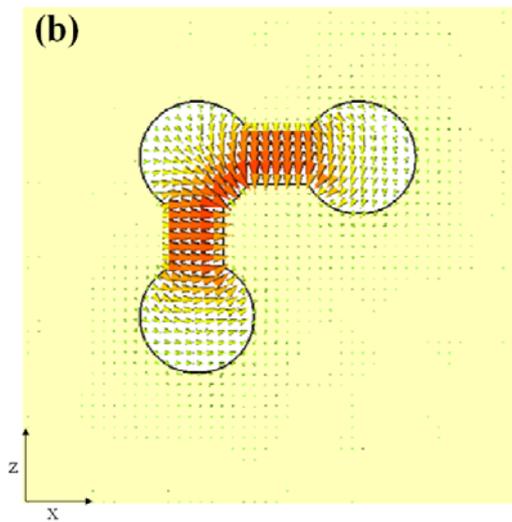

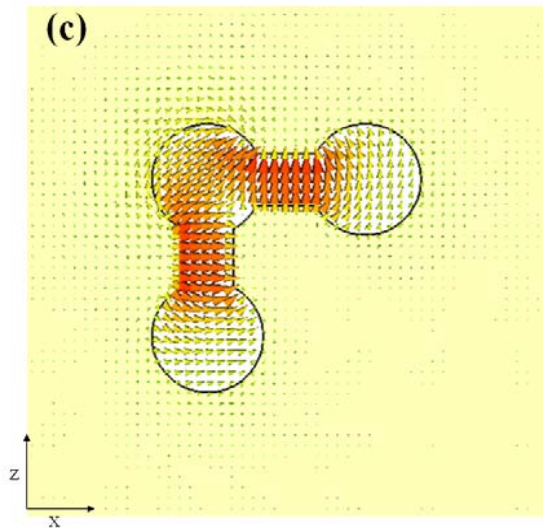

Fig.2 (a) Resonant frequencies. (b). Electric field for symmetry magnetic resonance mode (1550nm) (c). Electric field for asymmetry magnetic resonance mode (980nm).

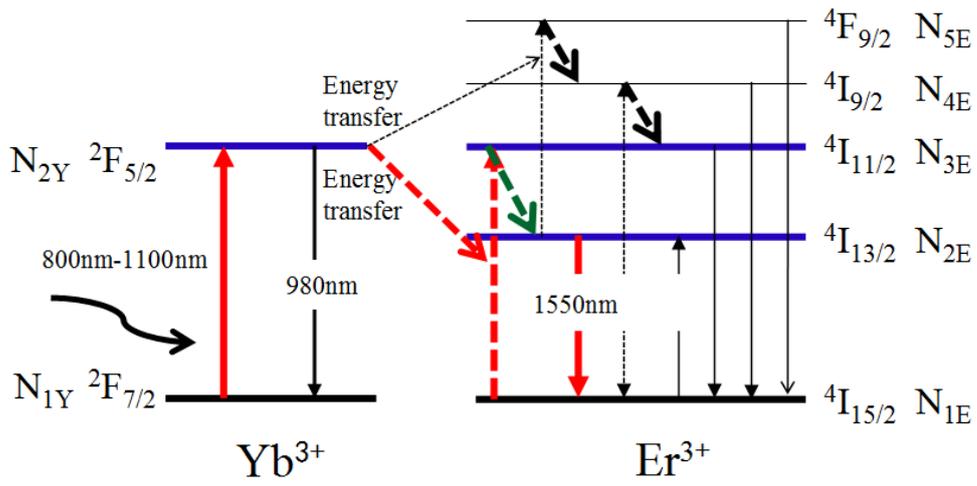

Fig.3 Energy level diagram for the Er–Yb co-doped system.